\documentclass{ws-procs975x65}

\begin{document}

\title{
MULTI-CHANNEL ALGEBRAIC SCATTERING THEORY\\
AND THE STRUCTURE OF EXOTIC COMPOUND NUCLEI}

\author{K. AMOS$^*$ and P. FRASER$^*$} 
\address{School of Physics, Melbourne University,\\
Victoria 3010, Australia\\
$^*$E-mail: amos@physics.unimelb.edu.au; pfraser@physics.unimelb.edu.au}

\author{D. van der KNIJFF$^*$}
\address{Advanced Computing Research, Information Division,\\
Melbourne University, Victoria 3010, Australia\\
$^*$E-mail: dirk@unimelb.edu.au}

\author{L. Canton$^*$ and G. Pisent$^*$}
\address{Istituto Nazionale di Fisica Nucleare, sezione di Padova, \\
e Dipartimento di Fisica dell'Universit$\grave {\rm a}$ di Padova,\\
via Marzolo 8, Padova I-35131\\
$^*$E-mail: luciano.canton@pd.infn.it; gualtiero.pisent@pd.infn.it}

\author{S. Karataglidis$^*$}
\address{ Department of Physics and Electronics, Rhodes University,\\
Grahamstown 6140, South Africa\\
$^*$E-mail: S.Karataglidis@ru.ac.za}

\author{J. P. Svenne$^*$}
\address{ Department of Physics and Astronomy, University of Manitoba,\\
and Winnipeg Institute for Theoretical Physics,\\
Winnipeg, Manitoba, Canada R3T 2N2\\
$^*$E-mail: svenne@physics.umanitoba.ca}

\begin{abstract}
A  Multi-Channel  Algebraic  Scattering (MCAS) theory is presented with
which  the   properties   of   a   compound  nucleus  are found from  a
coupled-channel problem. The method defines both the bound states  and
resonances of the compound nucleus,     even if the compound nucleus is
particle unstable. All resonances of the system are found no matter how 
weak and/or narrow.     Spectra of mass-7 nuclei and of ${}^{15}$F, and
MCAS results for a radiative capture cross section are presented.
\end{abstract}


\bodymatter

\section{Introduction}

            The MCAS method~\cite{Am03,Ca06}   has been formed to solve 
coupled-channel  Lippmann-Schwinger  equations  for  a chosen two-body 
system.   The theory is built upon sturmian expansions  of whatever one 
chooses  to  be  the matrix of interaction potential  functions for the 
system. The set of sturmians that form the basis of the  expansions are 
determined from the self-same interaction potentials and, if necessary,
can be selected to ensure that the Pauli principle is not violated. The
approach,  which is most suited to deal with low energies,  facilitates 
a  systematic  determination of all sub-threshold  bound and  resonance
states of the compound nucleus within any energy range considered. 

While the required starting  matrix  of   potentials may be constructed 
from any model of nuclear structure, to date we have used just a simple
collective rotational model to define those potentials with deformation 
taken to second order. Also, to date only low energy scattering from
light mass nuclei have been considered so that just the ground, first, 
and second excited states of the target nucleus have been used to define the 
channel couplings.

Isospin symmetry in mirror scattering systems,  save only for inclusion
of the Coulomb interaction,     has been used with MCAS to estimate the 
spectra of  nuclei  that  are  just  outside  of  the proton drip line.
$^{15}$F is one such case~\cite{Ca06a}.  The $A=7$ system also has been 
considered and, with a unique nuclear potential, the level structures of
$^7$He, $^7$Li, $^7$Be, and $^7$B have been obtained~\cite{Ca06}.    We 
have also used the scheme to consider low energy properties of 
${}^3$H-${}^4$He
and ${}^3$He-${}^4$He scattering to correlate  different  views  of the 
spectra of ${}^7$Li and ${}^7$Be.     Using the di-cluster structure of 
${}^7$Be found using MCAS,         the astrophysical $S$-factor for the 
radiative capture of ${}^3$He by ${}^4$He has been evaluated.

\section{The MCAS theory}
\label{mcasT}
In brief, the MCAS approach is based upon using sturmian functions as a
basis set to expand the chosen interaction potentials.  Each element in 
the interaction matrix then is a sum of separable interactions.     The
analytic properties of the $S$-matrix  from a separable Schr\"{o}dinger 
potential  gives  the  means  by which a full algebraic solution of the 
multichannel scattering problem can be realized.  As all details of the
MCAS theory have been published~\cite{Am03}, only salient features  are
repeated herein.    Consider a coupled-channel system for each allowed
scattering spin-parity $J^\pi$.    With the MCAS method, one solves the 
Lippmann-Schwinger (LS) integral equations in momentum space, i.e.
\begin{eqnarray}
T_{cc'}(p,q;E) = V_{cc'}(p,q)\, +
\frac{2\mu}{\hbar^2} \left[ \sum_{c'' = 1}^{\rm open} 
\int_0^\infty V_{cc''}(p,x) \frac{x^2}{k^2_{c''} - x^2 + i\epsilon} 
T_{c''c'}(x,q;E) \ dx \right.&&
\nonumber\\
\left.- \sum_{c'' = 1}^{\rm closed} \int_0^\infty V_{cc''}(p,x) 
\frac{x^2}{h^2_{c''} + x^2} T_{c''c'}(x,q;E) \ dx \right],\ &&
\label{multiT}
\end{eqnarray}
where the index  $c$  denotes  the  quantum  numbers that identify each 
channel uniquely.     Such requires specification of potential matrices 
$V_{cc'}^{(J^\pi)}(p,q)$.     The open and closed channels have channel 
wave numbers $k_c$ and $h_c$ for $E > \epsilon_c$  and $E < \epsilon_c$ 
respectively, and $\mu$  is the reduced mass.              Solutions of 
Eq.~(\ref{multiT}) are sought using expansions of the  potential matrix
elements in (finite) sums of energy-independent separable terms,
\begin{equation}
V_{cc'}(p,q) \sim  \sum^N_{n = 1} \chi_{cn}(p)\ \eta^{-1}_n\ 
\chi_{c'n}(q)\ .
\label{finiteS}
\end{equation}
The  key  to  the  method  is  the choice of the expansion form factors
$\chi_{cn}(q)$.          Optimal ones have been found from the sturmian
functions that are determined from the actual (coordinate space)  model
interaction       $V_{cc'}(r)$         initially chosen to describe the 
coupled-channel problem. The $\eta_n$ are energy scales associated with
the sturmian functions.  Details are given elsewhere~\cite{Am03}.

  Calculations  of the multichannel $T$- and scattering ($S$-) matrices 
involve a Green's function matrix,
\begin{eqnarray}
\left( {G}_0 \right)_{nn'} = 
\frac{2\mu}{\hbar^2}\left[ \sum_{c = 1}^{\rm open} 
\int_0^\infty \chi_{cn}(x) \frac{x^2}{k_c^2 - x^2 + i\epsilon} 
\chi_{cn'}(x)\ dx \right.\;\;\; &&
\nonumber\\
\left.  - \sum_{c = 1}^{\rm closed} \int_0^\infty 
\chi_{cn}(x) \frac{x^2}{h_c^2 + x^2} \chi_{cn'}(x)\ dx \right], &&
\label{xiGels}
\end{eqnarray}
The bound states of the compound system are defined 
by the zeros of matrix determinant for energy $E < 0$, namely
the zeros of    $\{ \left| \mbox{\boldmath $\eta$}-{\bf G}_0\right| \}$
when all channels in Eq.~(\ref{xiGels}) are closed. The $\mbox{\boldmath
$\eta$}$ is a  diagonal
matrix with entries $\left( \eta \right)_{nn'} = \eta_n\ \delta_{nn'}$. 

Elastic scattering observables are determined by the  on-shell   values 
($k_1 = k_1^\prime = k$) of the scattering matrices.    For the elastic
scattering of   neutrons  (spin $\frac{1}{2}$)  from  spin-zero targets 
$c = c'= 1$, and  $S_{11} \equiv S_{\ell}^J = S_{\ell}^{(\pm)}$ are
\cite{Am03}, 
\begin{equation}
S_{11} = 1 - i \pi \frac{2\mu k}{\hbar^2} \sum_{nn'=1}^M  \chi_{1n}(k) 
\frac{1}{\sqrt{\eta_n}} \left[\left({\bf 1} -  
\mbox{\boldmath $\eta$}^{-\frac{1}{2}}
{\bf G}_0\mbox{\boldmath $\eta$}^{-\frac{1}{2}}    
\right)^{-1}\right]_{nn'} \frac{1}{\sqrt{\eta_{n'}}} \chi_{1n'}(k).
\end{equation}
Diagonalization of the complex-symmetric matrix,
\begin{equation}
\sum_{n'=1}^N
{\eta_n}^{-\frac{1}{2}}\left[{\bf G}_0\right]_{nn'}
{\eta_{n'}}^{-\frac{1}{2}} \tilde{Q}_{n'r} = \zeta_r \tilde{Q}_{nr}\, ,
\label{eigen}
\end{equation}
establishes  the  evolution  of  the complex eigenvalues $\zeta_r$ with 
respect to energy. Resonant behavior occurs when an eigenvalue,
$\zeta_r$, as a function of $E$, crosses the unit circle in the Gauss 
plane near the point
(1,0).     It is evident that the elastic channel $S$ matrix has a pole 
structure  at  the  corresponding energy where the modulus of one of 
these eigenvalues approach unity, since  
\begin{equation}
\left[\left({\bf 1} -  \mbox{\boldmath $\eta$}^{-\frac{1}{2}}
{\bf G}_0\mbox{\boldmath $\eta$}^{-\frac{1}{2}} \right)^{-1}
\right]_{nn'}
=\sum_{r=1}^N \tilde{Q}_{nr}\frac{1}{1-\zeta_r}\tilde{Q}_{n'r} \, .
\end{equation}

\section{Results}

\subsection{The spectrum of ${}^{15}$F ($p$+${}^{14}$O)}
\begin{figure}[h]
\centering
\resizebox{0.8\columnwidth}{!}{\includegraphics{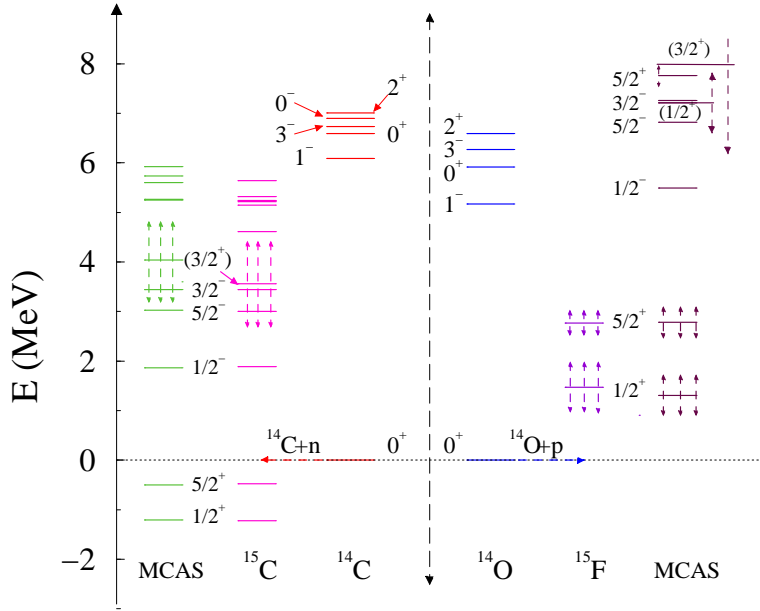}}
\caption{ Low energy spectra of ${}^{14,15}$C and of ${}^{14,15}$O, 
empirical~\cite{Aj91} and
from results of MCAS calculations.     The zero of energy is set to
that of the relevant mass-14 ground state.}
\label{Fig1}
\end{figure}

 Spectra, known and calculated (using MCAS~\cite{Ca06a}) for 
${}^{15}$C and ${}^{15}$F, are shown in Fig.~\ref{Fig1}.
The known properties of ${}^{15}$C agree well with MCAS results for the 
$n$+${}^{14}$C system.    Details have been specified fully in a recent 
paper~\cite{Ca06a}.  The spectrum of ${}^{15}$C has two bound states of
spin-parities $\frac{1}{2}^+$ (ground) and $\frac{5}{2}^+$.    They are
described dominantly as a single $sd$ shell neutron on the   ${}^{14}$C
ground state. Then there are three quite narrow resonances,  all having
negative parity, which lie within the spread of a broad $\frac{3}{2}^+$
resonant state.  That broad $\frac{3}{2}^+$ state was seen very clearly 
in  the   cross  section  from  a  measurement  of   ${}^{14}$C$(d,p)$.
It is noteworthy that there are no other bound states;    in particular 
none having negative parity. Such would occur if, in the $n$+${}^{14}$C
system,  the  $0p_{\frac{1}{2}}$  neutron orbit was  not Pauli blocked.
However, there are negative parity resonances, and  to find them in our
evaluations of ${}^{15}$C, required that the neutron $0p_{\frac{1}{2}}$
orbit only be Pauli hindered~\cite{Ca06a}. With the nuclear interaction
set, by including Coulomb interactions in MCAS evaluations,  properties
of the $p$+${}^{14}$O system (giving particle unstable ${}^{15}$F) were
determined. Those results are shown on the right of Fig.~\ref{Fig1}.
\begin{figure}[h]
\centering
\resizebox{0.8\columnwidth}{!}{\includegraphics{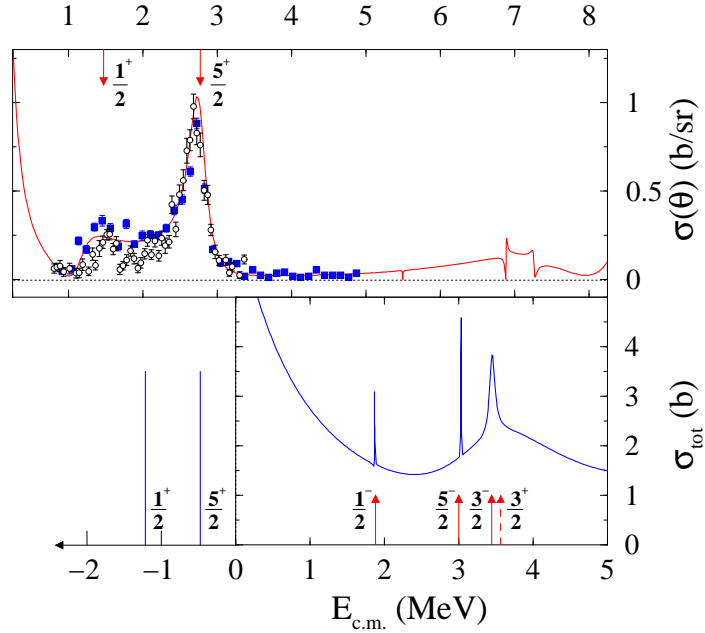}}
\caption{The elastic cross sections from scattering of  ${}^{14}$O ions 
from hydrogen at $\theta_{c.m.} = 180^\circ$ (top)  and  the  predicted 
(MCAS) total cross section for $n$-${}^{14}$C scattering (bottom).} 
\label{Fig2}
\end{figure}

    The two lowest resonance states of ${}^{15}$F have been observed in 
scattering cross-sections. Those data~\cite{Go04,Gu05}, and our results, 
are shown in Fig.~\ref{Fig2}.        In the top panel we show the cross 
sections from ${}^{14}$O scattering from hydrogen.      The MCAS result 
(solid curve) compares reasonably well with the data and it is as good as 
has been found with other analyses~\cite{Ca06a}. In the bottom panel of 
Fig.~\ref{Fig2} is a prediction of the total scattering  cross  section 
of neutrons from ${}^{14}$C for energies to 5 MeV.       There are four 
obvious resonances; three quite narrow (of negative parity)  and a very 
broad  $\frac{3}{2}^+$ one.   The three narrow resonances have partners
in the $p$+${}^{14}$O cross section.    That is shown in the top panel, 
while the broad $\frac{3}{2}^+$ structure      becomes more complex and
overlaps with two other states.   The zero of the energy scale has been 
placed to optimally match the $\frac{5}{2}^+$ bound state in ${}^{14}$C 
to the centroid of the analogous resonance state in ${}^{15}$F.

Analogues of the negative parity resonances in ${}^{15}$C are predicted 
in ${}^{15}$F.     But the origin of these new, narrow  negative-parity 
resonances in ${}^{15}$F differ from those of the   observed  low-lying
ones. They are compound resonances and are very affected by the   
Pauli-hindrance 
of the proton $0p_{\frac{1}{2}}$ orbit in the $0_2^+$ and $2^+$ excited 
states of ${}^{14}$O.

\subsection{Spectra of $^7$He, $^7$Li, $^7$Be, $^7$B}

  A single nuclear interaction matrix of potentials was found whose use 
with     appropriate Coulomb terms in MCAS calculations gave spectra of 
these nuclei in good agreement with known ones.              For $^7$Li 
($p$+${}^6$He) and $^7$Be ($n$+${}^6$Be),     the results are listed in 
Table~\ref{Table1}. 
\begin{table}
\tbl{Bound spectra of ${}^7$Li and of ${}^7$Be $^{\text a}$.} 
{\begin{tabular}{@{}ccc|c|cc|c@{}}
\toprule
$J^\pi$  & ${}^7$Li Exp. & $p+^6$He & $^3$H+$^4$He & ${}^7$Be Exp. & 
$n+^6$Be & $^3$He+$^4$He\\
\colrule
$\frac{3}{2}^-$ & $-$10.0      & $-$10.0 & $-$10.3 & $-$10.7      & 
$-$11.0 & $-$10.1 \\
$\frac{1}{2}^-$ & $-$9.5       & $-$9.5  & $-$9.74 & $-$10.2      & 
$-$10.7  & $-$9.88 \\
$\frac{7}{2}^-$ & $-$5.3 [0.06]& $-$5.3  & $-$5.4 [0.08] & 
$-$6.1 [0.18] & $-$6.4  & $-$6.0 [0.18] \\
$\frac{5}{2}^-$ & $-$3.4 [0.92] & $-$3.4  & $-$3.3 [0.83] & 
$-$4.0 [1.2] & $-$4.5  & $-$4.0 [1.19] \\
$\frac{5}{2}^-$ & $-$2.3 [0.08]& $-$0.3  & & $-$3.5 [0.44]& $-$1.6 & \\
$\frac{3}{2}^-$ & $-$1.2 [4.7] & $-$2.2  & & $--$         &   $--$ & \\
$\frac{1}{2}^-$ & $-$0.9 [2.8] & $-$0.9  & &              & $-$2.1 & \\
$\frac{7}{2}^-$ & $-$0.4 [0.44]& $-$0.4  & & $-$1.4 [?]   & $-$1.7 & \\
$\frac{3}{2}^-$ &  $--$        &    $--$ & & $-$0.8 [1.8] & $-$3.3 & \\
\botrule
\end{tabular}}
\begin{tabnote} 
$^{\text a}$ All energies are in reference to nucleon + mass 6 nucleus 
thresholds\\
\end{tabnote}
\label{Table1}
\end{table}
Therein the numbers in square brackets are the known experimental widths
obtained   from $t$+$\alpha$ ($^3$He +$\alpha$) reactions.  The nucleon
emission thresholds lie higher and so no nucleon widths arise from  the
MCAS calculations.  However, those evaluations produced 12 states to 15 MeV 
excitation in ${}^{7}$Li with the lowest 9 matching  known  spin-parity
states in the spectrum.     The next three calculated levels are in the
energy region in which two resonant states of undetermined  spin-parity
are known.        The five lowest lying states in the known spectrum of
${}^7$Be compare reasonably with the MCAS values. The calculations give 
more states than are known to date above an excitation energy of $\sim$
8.5 MeV, and there are a few crossings. 
\begin{figure}[ht]
\centering
\resizebox{0.8\columnwidth}{!}{\includegraphics{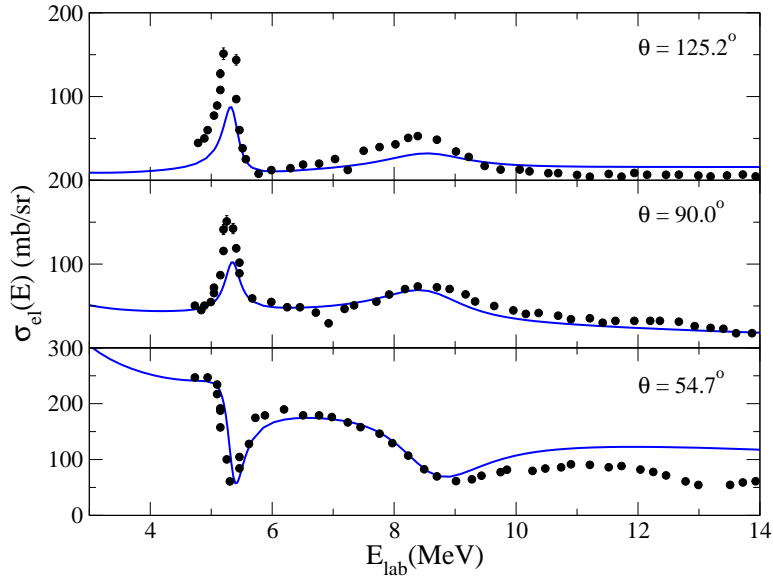}}
\caption{ Cross sections from $\alpha({}^3$He,${}^3$He$)\alpha$ 
scattering.} 
\label{Fig3}
\end{figure}

To consider other open reaction channels, we have used MCAS to evaluate
${}^3$H- and ${}^3$He-$\alpha$ scattering using a potential model  form
of a cluster system.      A nuclear interaction was found that gave the 
results listed in the fourth and seventh columns in Table~\ref{Table1}.
The calculated results agree well with the known values.  Note that the
energies are listed relative to the relevant nucleon+nucleus thresholds. 

    We used that interaction with MCAS to find elastic scattering cross 
sections at three center of mass scattering angles        ($54.7^\circ, 
90.0^\circ, 125.2^\circ$) at which data have been taken. The results of 
some of those calculations are shown by the             solid curves in 
Fig.~\ref{Fig3}.        The comparisons with data are very good, adding 
confirmation to our      definition of the basic di-cluster-interaction 
potential.

\subsection{MCAS and capture cross sections}
\begin{figure}[ht]
\centering
\resizebox{0.8\columnwidth}{!}{\includegraphics{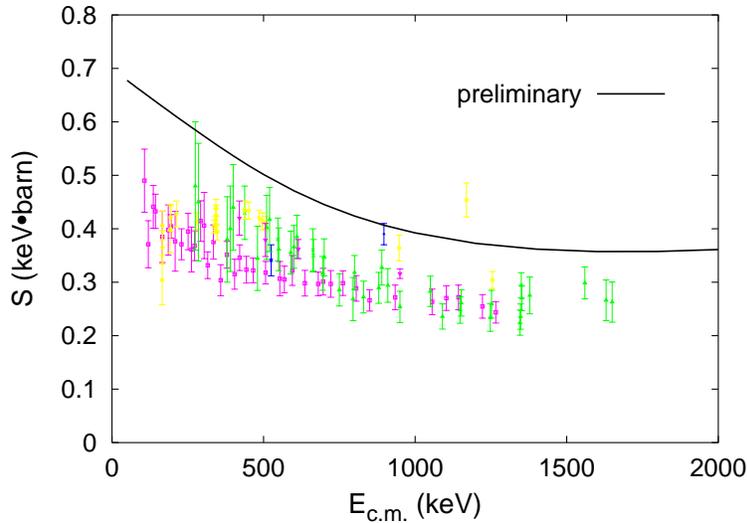}}
\caption{The calculated astrophysical $S$-factor from 
${}^3$He-${}^4$He capture compared with data~\cite{Be06}.}
\label{Fig4}
\end{figure}

      Recently, Canton, Levchuck, and Shebeko~\cite{Ca07} used the MCAS 
di-cluster model wave function,  though with an approximate  scattering 
wave function (hence the results are preliminary),      to evaluate the 
radiative capture cross section $\sigma_R(E)$        and from which the 
astrophysical $S$-factor is determined by 
\begin{equation}
S(E) = \sigma_R(E)\; E \exp{\{2\pi \eta(E)\}} .
\end{equation}
Therein $\eta(E)$ is the Sommerfeld parameter.       The results of the
calculated $S$-factor are compared    in    Fig.~\ref{Fig4}        with
data~\cite{Be06}.      The calculated $S$-factor overestimates the
measured data but by only 20\%.     There has been no arbitrary scaling
used, i.e. no asymptotic normalization or adjusted spectroscopic factor.
Given that  the  bound  state of ${}^7$Be is not expected to be 100\% a 
di-cluster, these results are most encouraging.           Of note, both
scattering and bound state functions can be given by MCAS so that there
is no  orthogonality problem as in past radiative capture calculations.

\section{Conclusions}
Applications of the MCAS theory have been made to discern the structure
of the compound nuclei underlying various di-cluster nuclear systems of
light mass.   Despite the simplicity of the collective model potentials
used, good agreement has been found with available data, even regarding
a nucleus lying beyond the proton drip line. Use of MCAS information to
evaluate radiative capture cross sections is a promising new 
development.

\section{Acknowledgements}
This research was supported
by the Italian MIUR-PRIN Project      ``Fisica Teorica del
Nucleo e dei Sistemi a Pi\`u Corpi'', by the Natural Sciences and
Engineering Research Council (NSERC), Canada, and by the National
Research Foundation of South Africa.


\bibliographystyle{ws-procs975x65}
\bibliography{ispun07-paper-amos}

\end{document}